\documentclass[sigconf]{acmart} % Use [sigconf,anonymous] for review
\usepackage{graphicx}
\graphicspath{{./figs/}}
\DeclareGraphicsExtensions{.png,PNG,.pdf,.PDF}

\usepackage[inline,shortlabels]{enumitem}
\setlist[itemize]{noitemsep,topsep=0pt,leftmargin=*}

\usepackage{cleveref} 
\crefname{section}{§\!}{§§\!}
\Crefname{section}{Section}{Sections}
\crefname{figure}{Fig.}{Figs.}
\Crefname{figure}{Figure}{Figures}
\crefname{table}{Table}{Tables}
\Crefname{table}{Table}{Tables}

\setcopyright{none}

\acmConference[e-Energy '26]{ACM International Conference on Future Energy Systems}{June 2026}{Banff, Canada}
\acmYear{2026}
\copyrightyear{2026}
\acmDOI{10.1145/3765611.3815430}
\acmISBN{979-8-4007-2199-1/2026/06}

\begin{document}

\title[Multi-market value-stacking with non-uniform FCR bidding]{Multi-market value-stacking: Battery control for combined imbalance participation and non-uniform FCR bidding}

%% STANDARD AUTHOR BLOCKS (TAPS COMPLIANT)
\author{Celle Hendrickx}
\affiliation{%
  \institution{Gent University - imec, IDLab}
  \city{Ghent}
  \country{Belgium}
}
\email{celle.hendrickx@ugent.be}

\author{Fabio Pavirani}
\affiliation{%
  \institution{Gent University - imec, IDLab}
  \city{Ghent}
  \country{Belgium}
}
\email{fabio.pavirani@ugent.be}

\author{Chris Develder}
\affiliation{%
  \institution{Gent University - imec, IDLab}
  \city{Ghent}
  \country{Belgium}
}
\email{chris.develder@ugent.be}

\renewcommand{\shortauthors}{Hendrickx, et al.}

\begin{abstract} 
The growing share of Renewable Energy Sources (RES) in modern power systems increases both grid imbalances and frequency deviations, reinforcing the need for ancillary services such as Frequency Containment Reserve (FCR) and passive balancing. Battery Energy Storage Systems (BESS) are well-suited for these services, but prior research typically relies on \emph{uniform} FCR bids that remain constant throughout the control period. Such static bids fail to fully exploit BESS flexibility, as they do not balance the trade-off between reserving energy for FCR delivery and using it for imbalance arbitrage, limiting the achievable value in value-stacking settings. To address this limitation, we propose a two-stage control framework for the European context that introduces \emph{non-uniform} FCR bids. In the first stage, we derive a time-varying bid sequence using data-driven Monte Carlo (MC) optimization. In the second stage, a Deep Reinforcement Learning (DRL) agent leverages the residual flexibility for real-time imbalance trading while proactively managing the State of Energy (SoE) to ensure compliance with FCR requirements. The framework is presented as a \emph{proof of concept}, highlighting the potential benefits of time-varying bidding strategies. By incorporating daily cycle budgets and time-varying reserve commitments, our approach achieves a 7.56\% profit increase compared to uniform baselines. These results show that non-uniform bidding can unlock additional value by more effectively aligning reserve obligations with rapidly changing imbalance opportunities. 
\end{abstract}

%% CCS Concepts



\maketitle
%================================
\section{Introduction}
%================================

The transition to carbon neutrality has reshaped power systems as variable RES displace conventional synchronous units. This shift increases frequency deviations and system imbalances due to RES volatility \cite{Kroposki2017, 4282051, WOO20113939}. Furthermore, as inverter-based RES lack the kinetic energy of traditional rotating masses, their penetration directly reduces system inertia \cite{ulbig2014impact}, necessitating faster frequency control mechanisms to maintain stability.

To address these challenges, European Transmission System Operators (TSOs) procure ancillary services, among which FCR acts as the system's first line of defense. BESS are particularly well suited for FCR provision due to their rapid response capabilities~\cite{4076076, 4282047, HU2022119512}, and their participation has grown steadily across the European synchronous area~\cite{baltputnis2024robust}. Declining lithium-ion battery costs further reinforce the economic viability of storage-based ancillary service provision~\cite{bnefLithiumIonBattery}.

However, BESS are considered Limited Energy Resources (LERs) and therefore require a real-time Energy Management Strategy (EMS) to ensure continuous availability for both upward and downward FCR activation~\cite{EliaDesignNoteFCR}. Early EMS approaches leveraged regulatory degrees of freedom, but increasing competition and stricter technical requirements have reduced their effectiveness~\cite{THIEN2017143, DRAHEIM2020101110, Baltputnis}, motivating a shift toward value-stacking, where BESS flexibility is coordinated across multiple services~\cite{tian2018stacked}. Success in this area is well-documented; multi-market studies have demonstrated significant gains through the coordination of energy and ancillary services \cite{spanish_virtual_storage}, dynamic application stacking \cite{englberger2020dynamic}, and joint bidding across reserve and spot markets \cite{seifert2024coordinated}.

The imbalance mechanism offers a natural solution for SoE management, providing a framework for both profitable arbitrage and systematic energy balancing \cite{engelhardt2022energy}. Recent research has confirmed the feasibility of managing SoE through intraday market participation \cite{zhang2025joint}, yet these methods rely on myopic control logic. By focusing strictly on immediate rewards, such strategies fail to capture the long-term value of the SoE, neglecting future opportunity costs and leaving significant revenue on the table. To mitigate this limitation, one may turn to traditional long-horizon optimization techniques such as Mixed-Integer Linear Programming (MILP) or Model Predictive Control (MPC), which optimize decisions over an explicit planning horizon. However, in fast-paced and highly stochastic FCR and imbalance environments, these methods are often constrained by their reliance on explicit models and accurate forecasts \cite{pavirani2024frequency, madahi2024distributional}. Consequently, DRL has emerged as a superior alternative. By deriving a learned value function, DRL agents implicitly internalize the long-term economic worth of the current SoE, effectively balancing immediate operational needs against future opportunity costs. This capability provides the foresight and adaptability required for effective real-time arbitrage and high-resolution operational control \cite{wang2018energy, madahi2024distributional, pavirani2024frequency}.

Despite advancements in DRL-based control, existing literature on joint FCR and imbalance participation largely assumes a \textit{uniform} bidding structure, where reserve capacity remains fixed across all time intervals \cite{pavirani2024frequency}. This forces the real-time controller to operate within a static, exogenously determined commitment. In contrast, \textit{non-uniform} bidding---which allows FCR commitments to vary across time blocks---has shown promise in deterministic settings, with reported profit increases of $\sim$10\% in day-ahead markets \cite{biermann2025multi}. However, it remains unclear if these benefits persist under stochastic, real-time imbalance dynamics. Together, these observations expose two concurrent gaps: existing non-uniform bidding frameworks do not address real-time stochastic control, while DRL-based approaches that handle such dynamics have considered only uniform bids. This motivates our central research question: \textit{can we unlock additional revenue by adopting non-uniform FCR bids in combination with high-resolution, real-time imbalance control?}

We address this through a two-stage control framework. In the first stage, a data-driven MC procedure derives a time-varying non-uniform FCR bid sequence using ex-post information for bid selection, without imposing perfect-foresight assumptions on the real-time control layer. In the second stage, a DRL controller exploits the residual flexibility for real-time imbalance trading while maintaining SoE feasibility under a daily cycle budget. The bidding layer thus actively shapes a time-varying feasibility envelope that the real-time controller exploits under stochastic activation and price dynamics.

The remainder of this paper is organized as follows. \Cref{sec:problem} outlines the overall problem setting. \Cref{sec:methodology} presents the proposed two-stage control framework. \Cref{sec:exp_setup} describes the experimental setup, \Cref{sec:results_discussion} reports and interprets the results, and \Cref{sec:conclusions} concludes the paper.

%================================
\section{Problem Formulation}
\label{sec:problem}
%================================

\subsection{Battery Model}
\label{sec:battery_model}
We consider a BESS characterized by its nominal power $P_{\text{nom}}$ in MW, energy capacity $E_{\text{cap}}$ in MWh, and charging/discharging efficiencies $\eta_c, \eta_d \in [0, 1]$. To model bidirectional operation, the total battery power $P_t^{\text{total}}$ (\cref{eq:total_power}) is decomposed into charging and discharging components $P_t^{+}, P_t^{-} \in \mathbb{R}^+$. The SoE evolves according to \cref{eq:soe_update}, with $P_t^{+}, P_t^{-} \in [0, P_{\text{nom}}]$:
\begin{equation}\label{eq:soe_update}
    E_{t+1} = E_t + \left( \eta_c P_t^{+} - \frac{P_t^{-}}{\eta_d} \right)\Delta t.
\end{equation}

To account for battery lifetime considerations, we measure the cumulative discharge throughput $C$ over time (\cref{eq:cycle_limit}). This quantity is later used in the control framework (\cref{sec:DRL}) to penalize excessive cycling through the DRL reward design.
\begin{equation}\label{eq:cycle_limit}
    C = \sum_{k=t-H}^{t} \frac{P_k^{-}\,\Delta t}{E_{\text{cap}}},
    \qquad H = 1440~\text{min}.
\end{equation}

\subsection{FCR Requirements}
\label{sec:fcr_market}
In the European FCR framework, an accepted bid $P_b^{\textsc{fcr}}$ requires the provider to continuously adjust its power output according to real-time frequency deviations at a second-level resolution. The required activation $P_t^{\textsc{fcr}}$ is determined by a fixed, standardized rule that specifies how much power must be delivered for a given frequency deviation \cite{EliaDesignNoteFCR}.

For LERs such as BESS units, the provider must ensure sufficient SoE to sustain a full activation in either direction for a fixed duration $T_{\text{res}}$, which imposes the operational SoE margin:
\begin{equation}\label{eq:soe_margin}
    P_b^{\textsc{fcr}} \> T_{\text{res}}
    \;\le\;
    E_t
    \;\le\;
    E_{\text{cap}} - P_b^{\textsc{fcr}} \> T_{\text{res}}.
\end{equation}
Non‑compliance with these limits is considered non‑delivery and can result in penalties, temporary suspension, or exclusion from future FCR participation~\cite{EliaDesignNoteFCR}. As a result, each FCR bid directly constrains the usable energy band and determines how much flexibility remains for other services such as imbalance arbitrage.

\subsection{Imbalance Settlement}
\label{sec:imbalance_market}
The imbalance settlement mechanism provides both a revenue opportunity and a means for SoE regulation. At each control step (detailed in \cref{sec:DRL}), the operator selects an imbalance power setpoint $P_t^{\text{imb}}$ based on price signals and system state. Simultaneously, the asset delivers the FCR activation $P_t^{\textsc{fcr}}$ driven by real-time frequency deviations. The total power injection is thus:
\begin{equation}\label{eq:total_power}
    P_t^{\text{total}} = P_t^{\textsc{fcr}} + P_t^{\text{imb}}.
\end{equation}

%================================
\section{Methodology}
\label{sec:methodology}
%================================
We use a two-stage approach to decouple long-horizon FCR bidding decisions from fast real-time BESS control: Stage~1 determines the FCR bid sequence using a data-driven evaluation of candidate bids (\cref{sec:fcr_bid}), while Stage~2 uses a DRL controller to operate the BESS under these commitments (\cref{sec:DRL}). This allows us to treat long-horizon bidding and fast real‑time control in a coordinated yet modular way.

%------------------------------------------
\subsection{Data-driven FCR Bid Optimization}
\label{sec:fcr_bid}
%------------------------------------------

For each 4-hour block $b$, Stage~1 evaluates a discrete set of candidate FCR bids to identify the time-varying bid sequence with the highest expected profit. The procedure is carried out \emph{ex post} using realized frequency and price data, providing an upper-bound benchmark for the potential of non-uniform bidding without relying on forecasting. The resulting bids define the time-varying reserve commitments (\cref{eq:soe_margin}) under which the real-time controller (Stage~2) operates.

For each block, we consider bids at 1\,MW granularity,
\[
P_{b,k}^{\textsc{fcr}} \in \{0, 1, \dots, P_{\text{nom}} - 1\},
\]
reflecting the minimum bidding resolution imposed by European FCR procurement rules \cite{EliaDesignNoteFCR}. The nominal rating $P_{\text{nom}}$ is excluded to preserve residual flexibility for SoE management through passive balancing. To account for uncertainty in the initial SoE at the start of each block, we run $50$ MC simulations per candidate bid, drawing initial SoE values across the full feasible reserve band, including boundary points for compliance verification.

Each MC simulation covers the full 4-hour block at 1\,s resolution and incorporates:
\begin{itemize}
    \item \textbf{FCR activation:} second-level power response computed from historical frequency data;
    \item \textbf{Financial settlement:} FCR capacity remuneration and 15\,min imbalance settlement.
\end{itemize}
Two challenges arise in this setup. 
First, the simulation requires an SoE-management strategy to prevent violations of the FCR constraints while enabling imbalance revenues. We address this by using a simple \emph{Heuristic Imbalance Controller} that applies rule-based corrective actions derived from SoE zones and price-percentile triggers. As this heuristic differs from the DRL controller used in Stage~2 (\cref{sec:DRL}), a mild policy mismatch is introduced; however, its impact on the bid selection is likely limited, since all candidate bids are evaluated under the same heuristic policy. 
Second, different bids can lead to different end-of-block SoE levels, introducing bias because an artificially depleted battery provides less usable energy in the next block and thus incurs an implicit future cost. Hence, to ensure consistency, we include a terminal-value adjustment in the profit metric:
\begin{equation}
    J_{b,k}^{\mathrm{adj}}
    = R_{b,k}^{\textsc{fcr}}
    + \Pi_{b,k}^{\text{imb}}
    + \bar{\pi}_{b+1}\,\Delta E_{b,k},
\end{equation}
where $R_{\textsc{fcr}}$ is the FCR capacity revenue, $\Pi_{\text{imb}}$ the imbalance profit, $\bar{\pi}_{k+1}$ the median imbalance price of the following block, and $\Delta E$ denotes the change in SoE between the start and end of the block.
For each block, we choose the candidate bid with the highest expected adjusted profit,
\begin{equation}
    P_{b}^{\textsc{fcr}*}
    = \arg\max_{P_{b,k}^{\textsc{fcr}}}
      \mathbb{E}\!\left[J_{b,k}^{\mathrm{adj}}\right].
\end{equation}

%-------------------------------
\subsection{RL-Based Imbalance Control}
\label{sec:DRL}
%-------------------------------

The real-time imbalance trading problem follows a Markov Decision Process (MDP) framework, using a Double Deep Q-Network (DDQN) \cite{DDQN} as the underlying agent. The agent operates at a 1-minute decision interval $\Delta t_{\text{imb}}$, aligning with the update frequency of imbalance price signals published by Elia, the Belgian TSO. This high-frequency interval ensures the strategy remains compatible with balancing market dynamics across various European synchronous areas, preserving the generality of the approach while maintaining consistency with the data used in our experimental evaluation (\cref{sec:experiments}).

The agent receives a normalized state vector $\mathbf{s}_t$ that aggregates three categories of information:
\begin{itemize}
    \item \textbf{Market Signals}: 1-minute imbalance price indicators $\hat{\pi}_t$ together with time features such as the quarter-hour, minute within the quarter-hour, and month.
    \item \textbf{Physical State}: the current SoE and the cumulative cycle-usage (\cref{eq:cycle_limit}) ratio relative to the daily budget $C_{\text{max}}$.
    \item \textbf{Constraint Context}: the available operational headroom, quantified by the distance between the current SoE and both active and upcoming SoE bounds (\cref{eq:soe_margin}), as well as the current and next FCR commitments $P_{b,k}^{\textsc{fcr}*}$ and the time remaining until the next 4-hour block.
\end{itemize}
The action space $\mathcal{A}$ consists of discrete \emph{charge}, \emph{idle}, and \emph{discharge} decisions, mapped to the available residual flexibility $P^{\text{res}} = P_{\text{nom}} - P_{b}^{\textsc{fcr}*}$. This ensures that the total power output $P^{\textsc{fcr}} + P^{\text{imb}}$ remains within the converter limit $P_{\text{nom}}$ at all times. Operational safety is enforced via a two-tier mechanism: 
\begin{enumerate*}[(i)]
    \item an \emph{action mask} that filters out imbalance setpoints likely to violate SoE constraints within the upcoming interval, and
    \item a \emph{reactive corrective override} that adjusts actions if the SoE is pushed outside of the FCR bounds (\cref{eq:soe_margin}) by FCR activations caused by stochastic, sub-minute frequency deviations.
\end{enumerate*}

To balance profitability with asset health and operational robustness, the agent maximizes a multi-objective reward function $r_t$ reflecting three priorities: (i)~maximizing net imbalance revenue, (ii)~maintaining operational safety, and (iii)~preserving battery health. The resulting reward function is:
\begin{equation}
    r_t = r_t^{\text{imb}} + r_t^{\text{soe}} + r_t^{\text{cycle}} + r_t^{\text{override}}.
\end{equation}
Here, $r_t^{\text{imb}}$ represents the imbalance revenue contribution, while $r_t^{\text{soe}}$ penalizes both the proximity to and the violation of SoE limits. The term $r_t^{\text{override}}$ accounts for instances where the safety layer intervenes to enforce operational bounds. Cycle aging is incorporated through a soft penalty, $r_t^{\text{cycle}}$, applied only when the daily throughput budget is exceeded:
\begin{equation}
    r_t^{\text{cycle}} = -\lambda_c \cdot \max\!\left(0,\, C - C_{\text{max}}\right).
\end{equation}

While prior work has applied Soft Actor-Critic (SAC) in imbalance control \cite{pavirani2024frequency, madahi2024distributional}, we use DDQN as a stable and well-established baseline. Since our goal is to evaluate the impact of non-uniform FCR bidding rather than to benchmark RL algorithms, DDQN provides an adequate and transparent control layer for this study. The controller architecture remains compatible with more advanced agents, making their integration an interesting direction for future research.

%================================
\section{Experiments and Results}
\label{sec:experiments}
%================================

In this section, we evaluate the performance of the proposed two-stage framework. We first describe the experimental setup and data sources, followed by the training protocol, and evaluation metrics used to assess the value-stacking performance of the non-uniform strategy against its uniform counterparts (\cref{sec:exp_setup}). Finally, we discuss the results in \cref{sec:results_discussion}.

%-------------------------------------------------------------------------------
\subsection{Experimental Setup and Data Sources}
\label{sec:exp_setup}
%-------------------------------------------------------------------------------

Following prior work~\cite{pavirani2024frequency}, we evaluate our model in the Belgian context using a grid-scale BESS ($P_{\text{nom}} = 10$\,MW, $E_{\text{cap}} = 20$\,MWh, $\eta_c = \eta_d = 90\%$). To ensure longevity, we limit operation to 420 equivalent full cycles per year ($C_{\text{max}} \approx 1.15$ daily).

Simulations use historical data for the full year 2022, integrating high-resolution frequency and market signals. FCR activation is modeled using 1\,s frequency measurements from the Synchronous Area of Continental Europe (Netztransparenz~\cite{Netztransparenz}), applying the Elia activation rules~\cite{EliaDesignNoteFCR}; in the Belgian context, this sets $T_{\text{res}} = 25$\,min (\cref{eq:soe_margin}). Capacity revenue is computed using 4-hour FCR clearing prices from the Regelleistung portal~\cite{regelleistung_fcr}, while imbalance participation uses 1-minute price forecasts and 15-minute settlement prices from the Elia Open Data platform~\cite{elia_imbalance_prices}.

To ensure robustness and avoid temporal data leakage, we use a chronological split: the first 20 days of each month for training, the next 5 for validation, and the remaining days for out-of-sample testing.

We compare two bidding paradigms:
\begin{itemize}
    \item \textbf{Uniform Baselines:} Ten agents, each corresponding to a fixed FCR bid $P^{\textsc{fcr}} \in \{0,\dots,9\}$~MW, following~\cite{pavirani2024frequency}. The 10~MW case is excluded due to the absence of remaining flexibility for SoE management.
    \item \textbf{Non-Uniform Agent:} A single agent trained under the time-varying bid sequence $P_b^{\textsc{fcr}*}$ obtained from Stage~1 (\cref{sec:fcr_bid}), requiring adaptation to block-dependent SoE margins.
\end{itemize}
All agents share the same DDQN architecture and are trained using standard hyperparameters, with model selection based on peak validation profit.

Beyond total annual profit, we assess 
\begin{enumerate*}[(i)]
    \item profit decomposition (FCR revenue vs.\ imbalance arbitrage), and 
    \item asset health via the total number of cycles used.
\end{enumerate*}

%===============================================================================
\subsection{Results and Discussion}
\label{sec:results_discussion}
%===============================================================================
The experimental results demonstrate that the DRL agent successfully learns a profitable multi-market policy while strictly adhering to the specified operational constraints. The economic performance of our uniform bidding agents aligns closely with the benchmarks established in~\cite{pavirani2024frequency}, showing a peak between the 5\,MW and 8\,MW bids, thereby validating the fidelity of our simulation environment.

\begin{figure}[!t]
    \centering
    \includegraphics[width=\linewidth]{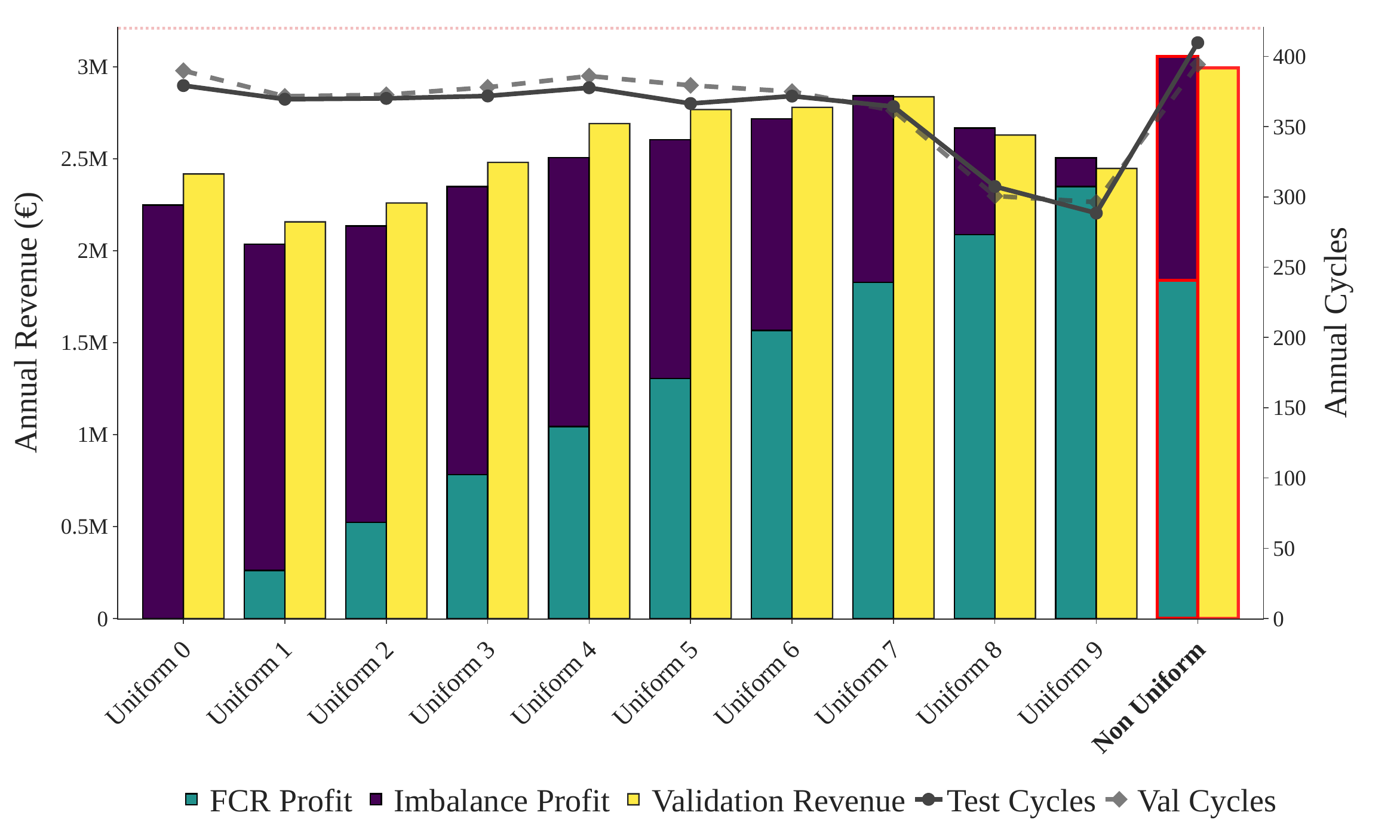}
    \Description{Bar chart comparing yearly revenues and cycle usage for uniform and non-uniform FCR bidding strategies. Each bar represents the total annual revenue achieved under a fixed uniform FCR bid between 0 and 9 MW, alongside the corresponding yearly cycle throughput. The non-uniform strategy achieves the highest total net revenue while keeping cycling within the daily budget, outperforming all uniform bidding baselines.}
    \caption{Yearly test and validation revenues and cycles of uniform and non-uniform bidding strategies}
    \vspace{-12pt} % Pulls the NEXT figure up closer to this one
    \label{fig:results}
\end{figure}

As shown in \Cref{fig:results}, the proposed non-uniform bidding strategy achieves a total net profit of 3,06\,M~EUR, outperforming the best-performing uniform baseline (2,84\,M~EUR). This represents a 7.56\% increase in total net profit. Interestingly, the FCR-specific revenue remains comparable between the non-uniform and the best-performing uniform strategies, suggesting that the first-stage optimizer intelligently allocates capacity during periods of peak FCR prices to maintain a strong baseline income. 

The primary performance gain, however, stems from the increased flexibility afforded to the real-time controller. By dynamically adjusting FCR commitments, the non-uniform strategy strategically expands the operational SoE margins and releases additional power headroom during volatile imbalance windows. This allows the DRL agent to execute more assertive imbalance arbitrage that would otherwise be restricted under a static FCR bid. Consequently, the first-stage optimizer acts as an enabler for the real-time controller; by selectively reducing FCR obligations during periods of high-margin imbalance opportunities, the system maximizes total market capture without sacrificing the baseline revenue typically associated with FCR participation. While this strategy results in a higher number of charging cycles compared to uniform baselines (see the line in \cref{fig:results}), the agent successfully keeps the total throughput within the prescribed daily cycle limit $C_{\text{max}}$, demonstrating a more effective utilization of the available asset headroom.

\begin{figure}[!t]
    \centering
    \includegraphics[width=\linewidth]{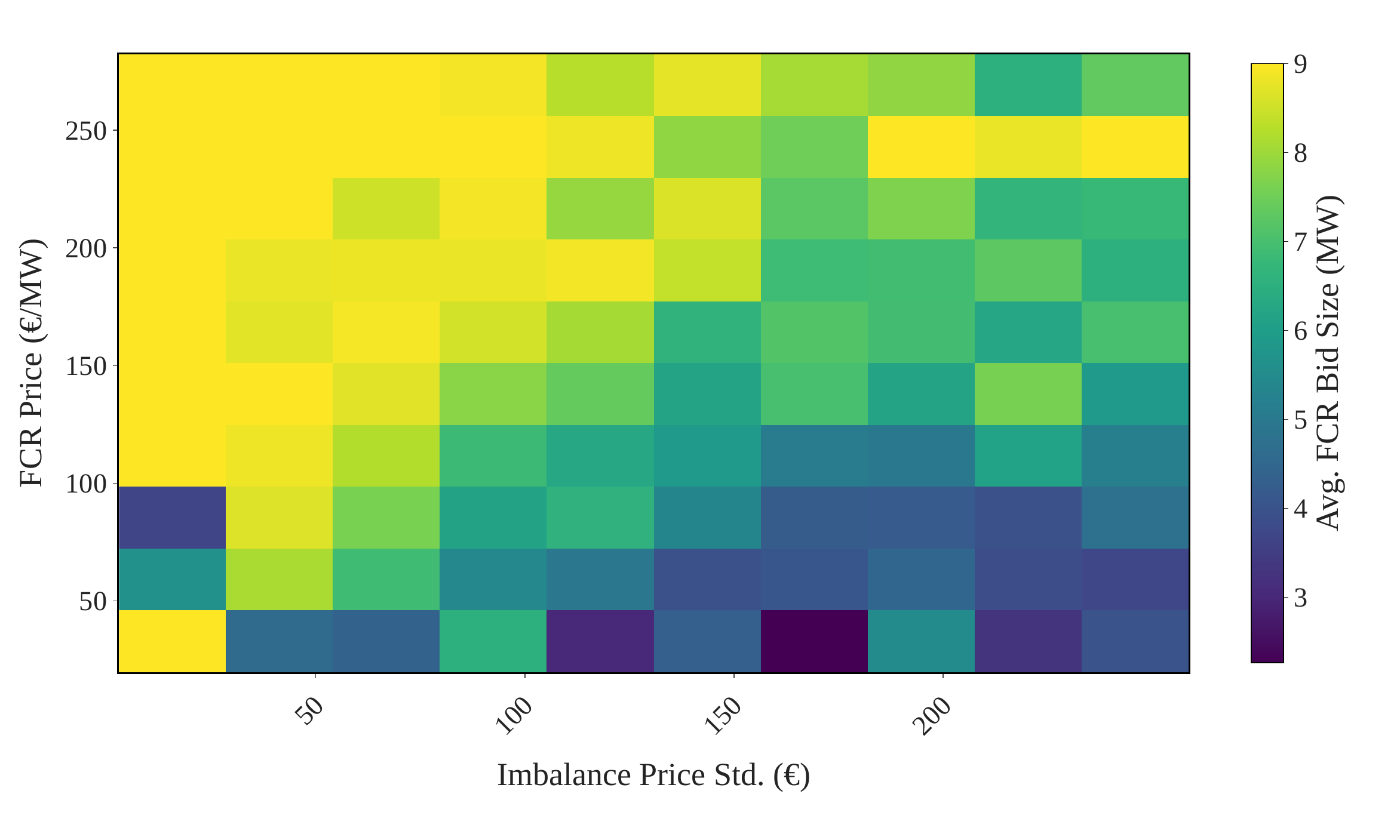}
    \Description{Heatmap showing the average FCR bid size selected by the first-stage optimizer as a function of the FCR price (vertical axis) and the standard deviation of the imbalance price (horizontal axis). Higher FCR prices correspond to larger bids, while higher imbalance-price volatility leads to smaller bids. The heatmap exhibits a roughly diagonal transition boundary, with capacity allocated to FCR in high-price, low-volatility regions and shifted toward imbalance participation when volatility is high or FCR remuneration is low.}
    \caption{Average FCR bid size (MW) relative to FCR price and imbalance volatility ($\sigma$) for each FCR block.}
    \vspace{-12pt} % THIS pulls the "Conclusion" text UP toward the figure
    \label{fig:strategy_heatmap}
\end{figure}

This bidding behavior is further illustrated in \cref{fig:strategy_heatmap}. The first-stage optimizer consistently selects larger FCR bids when FCR prices are high, while increasing imbalance-price volatility leads to smaller bids as more flexibility is reserved for real-time control. The resulting pattern exhibits a roughly diagonal transition: capacity is allocated to FCR in high-price, low-volatility regions, and shifts toward imbalance arbitrage as volatility rises or FCR remuneration weakens.

Beyond the specific FCR–imbalance setting, the broader structure of our design can naturally generalize to other value‑stacking combinations. Because the RL controller operates under time‑varying SoE feasibility windows induced by block‑dependent market commitments and stochastic activations, the same control logic applies to services such as aFRR, hybrid PV–BESS smoothing, or DA/ID scheduling combined with ancillary products. All of these services impose product‑specific, time‑dependent availability constraints analogous to the FCR 25‑minute rule and require an EMS capable of ensuring compliance while exploiting short‑term price opportunities. The proposed two‑stage ``bid then control residual flexibility’’ architecture therefore provides a reusable abstraction for such settings, with extensions primarily requiring parameterization of activation profiles, rather than changes to the underlying control methodology.

\vspace{-5pt} 
%================================
\section{Conclusion}
\label{sec:conclusions}
%================================
This paper presented a two-stage framework for joint FCR and imbalance participation that moves beyond the conventional assumption of uniform, static bidding. By combining
\begin{enumerate*}[(1)]
\item\label{it:stage1} a data-driven %Stage~1 
optimizer for non-uniform FCR profiles with 
\item a DRL-based real-time controller, our approach achieves a 7.56\% increase in annual profit over the best uniform baseline, for our case study in the Belgian market.
\end{enumerate*}
These gains stem from the enhanced operational flexibility unlocked through time-varying FCR commitments. Specifically, non-uniform bidding widens the feasible SoE margins and increases power headroom during periods of high imbalance volatility, enabling the controller to capitalize on price spikes while maintaining full FCR compliance and adhering to daily cycling limits. While these results demonstrate the structural value of non-uniform bidding, this study uses ex-post evaluations in stage~\ref{it:stage1}. A natural progression for future work involves integrating proactive, forecasting-based bid selection and investigating more robust RL architectures beyond DDQN. More broadly, the ``bid then control residual flexibility'' paradigm is applicable to other value-stacking contexts with time-dependent constraints, such as aFRR, or hybrid PV--BESS systems. As ancillary service markets evolve, such coordinated bidding and control strategies offer a promising pathway for enhancing the economic viability of energy storage assets.

\begin{acks}
This research was partly funded by the Flemish Government through the ``Onderzoeksprogramma Artifici\"{e}le Intelligentie (AI) Vlaanderen'' programme.
\end{acks}

\balance
%\clearpage
\bibliographystyle{ACM-Reference-Format}
\bibliography{references}

\end{document}